\documentclass[twocolumn]{aastex63}
\usepackage{amsmath}
\usepackage{hyperref}
\usepackage{graphicx}
\usepackage{booktabs}
\usepackage{diagbox}

\usepackage{xcolor}
\usepackage{bold-extra}

\shorttitle{H$\alpha$ LF from GLASS-JWST}
\shortauthors{Pang et al.}

\begin{document}

\title{The GLASS-JWST Early Release Science Program. V. H$\alpha$ luminosity functions at $z\sim1.3$ and $z\sim2.0$}

\author[0009-0005-3823-9302]{Yuxuan Pang}
\affiliation{School of Astronomy and Space Science, University of Chinese Academy of Sciences (UCAS), Beijing 100049, China}

\author[0000-0002-9373-3865]{Xin Wang}
\affiliation{School of Astronomy and Space Science, University of Chinese Academy of Sciences (UCAS), Beijing 100049, China}
\affiliation{National Astronomical Observatories, Chinese Academy of Sciences, Beijing 100101, China}
\affiliation{Institute for Frontiers in Astronomy and Astrophysics, Beijing Normal University, Beijing 102206, China}

\author[0000-0002-8460-0390]{Tommaso Treu}
\affiliation{Department of Physics and Astronomy, University of California, Los Angeles, 430 Portola Plaza, Los Angeles, CA 90095, USA}

\author[0009-0006-1255-9567]{Qianqiao Zhou}
\affiliation{School of Astronomy and Space Science, University of Chinese Academy of Sciences (UCAS), Beijing 100049, China}

\author[0009-0007-6655-366X]{Shengzhe Wang}
\affiliation{National Astronomical Observatories, Chinese Academy of Sciences, Beijing 100101, China}
\affiliation{School of Astronomy and Space Science, University of Chinese Academy of Sciences (UCAS), Beijing 100049, China}

\author[0000-0002-7350-6913]{Xue-Bing Wu}
\affiliation{Department of Astronomy, School of Physics, Peking University, Beijing 100871, People's Republic of China}
\affiliation{Kavli Institute for Astronomy and Astrophysics, Peking University, Beijing 100871, People's Republic of China}

\author[0000-0001-5984-03925]{Maru\v{s}a Brada{\v c}}
\affiliation{University of Ljubljana, Faculty of Mathematics and Physics, Jadranska ulica 19, SI-1000 Ljubljana, Slovenia}
\affiliation{Department of Physics and Astronomy, University of California Davis, 1 Shields Avenue, Davis, CA 95616, USA}

\author[0000-0002-3254-9044]{Karl Glazebrook}
\affiliation{Centre for Astrophysics and Supercomputing, Swinburne University of Technology, PO Box 218, Hawthorn, VIC 3122, Australia}

\author[0000-0003-4570-3159]{Nicha Leethochawalit}
\affiliation{National Astronomical Research Institute of Thailand (NARIT), Mae Rim, Chiang Mai, 50180, Thailand}

\author[0000-0001-6919-1237]{Matthew A. Malkan}
\affiliation{Department of Physics and Astronomy, University of California, Los Angeles, 430 Portola Plaza, Los Angeles, CA 90095, USA}

\author[0000-0003-2804-0648]{Themiya Nanayakkara}
\affiliation{Centre for Astrophysics and Supercomputing, Swinburne University of Technology, Hawthorn, VIC 3122, Australia}

\author[0000-0003-0980-1499]{Benedetta Vulcani}
\affiliation{INAF, Osservatorio Astronomico di Padova, Vicolo dell'Osservatorio 5, 35122 Padova, Italy}

\author[0000-0003-3108-0624]{Peter J. Watson}
\affiliation{INAF, Osservatorio Astronomico di Padova, Vicolo dell'Osservatorio 5, 35122 Padova, Italy}

\author[0000-0003-1718-6481]{Hu Zhan}
\affil{National Astronomical Observatories, Chinese Academy of Sciences, Beijing 100101, China}
\affil{Kavli Institute for Astronomy and Astrophysics, Peking University, Beijing 100871, China}


\correspondingauthor{Yuxuan Pang, Xin Wang}
\email{pangyuxuan@ucas.ac.cn, xwang@ucas.ac.cn}

\begin{abstract}
We present H$\alpha$ luminosity function (LF) measurements at redshifts $z\sim1.3$ and $z\sim2.0$ using JWST NIRISS grism data from the GLASS-JWST survey. Based on emission lines spectroscopically identified in the F115W, 
F150W and F200W filters, we select 99 H$\alpha$ emitters. Through detailed effective volume and completeness analysis for each source, we construct the H$\alpha$ LF in two redshift bins. Thanks to the sensitivity of NIRISS WFSS and gravitational lensing magnification, our sample reaches intrinsic H$\alpha$ luminosities $\sim$10 times deeper than previous grism surveys, down to $L_{\rm H\alpha}\sim10^{40.5}~\rm erg~s^{-1}$ at $z\sim1.3$ and $L_{\rm H\alpha}\sim10^{40.9}~\rm erg~s^{-1}$ at $z\sim2.0$ with completeness larger than 0.8, corresponding to star formation rates of 0.4 and 1.0 $M_{\odot}~\rm yr^{-1}$, respectively. We robustly constrain the faint-end slope of the H$\alpha$ luminosity function to be $-1.50^{+0.14}_{-0.08}$ at $z\sim1.3$ and $-1.60^{+0.17}_{-0.09}$ at $z\sim2.0$ after considering the cosmic variance of $\sim 20\%$, consistent with previous estimations. The emission-line samples presented here will enable further detailed studies of galaxy properties including metallicities.
We find a negligible contribution from bright active galactic nuclei in our sample. We estimate integrated cosmic star formation rate densities of $0.097^{+0.015}_{-0.016}~M_{\odot}~\rm yr^{-1}~Mpc^{-3}$ at $z\sim1.3$ and $0.129^{+0.025}_{-0.030}~M_{\odot}~\rm yr^{-1}~Mpc^{-3}$ at $z\sim2.0$. 
The methodology presented here can be readily applicable to other JWST slitless spectroscopic datasets and future wide-field slitless surveys, including those from Euclid, Roman, and the Chinese Space Station Telescope.

\end{abstract}


\section{Introduction} \label{sec: intro}
The Cosmic Noon era between $z\sim3$ and $z\sim1$ was one of the most pivotal phases in the universe's history: half of the stellar mass observed in galaxies today was formed over 3.5 Gyr,  with the cosmic star-formation rate (SFR) volume density peaking around $z\sim2$ \citep{2008ApJ...675..234P, 2010ApJ...709.1018V, 2011A&A...528A..35M, 2013A&A...556A..55I, 2013ApJ...777...18M, 2014ARA&A..52..415M, 2020ARA&A..58..661F}. Back to $z\sim3$, star-forming galaxies still follow the main sequence (MS) between SFR and stellar mass through the redshift evolution
\citep{2011ApJ...739L..40R, 2014ApJS..214...15S}, as well as other scaling relations involving size, kinematics, and metal and gas content \citep{2014ApJ...788...28V, 2017ApJ...842..121U, 2019A&ARv..27....3M, 2020ARA&A..58..157T}. The comoving accretion rate of supermassive black holes (SMBHs) follows a similar evolutionary trend, consistent with the coevolution of central black holes and their host galaxies \citep{2006MNRAS.368.1395M, 2014ApJ...786..104U, 2014MNRAS.439.2736D, 2015MNRAS.451.1892A}. 

Measuring the cosmic star formation rate--a key to galaxy evolution--across the universe's history remains a major goal in observational cosmology. Therefore, multiple SFR indicators have been developed and adopted \citep[see reviews by][]{2012ARA&A..50..531K, 2014ARA&A..52..415M}. Among them, the H$\alpha$ emission line has been most widely used as one of the most reliable nebular-line tracers of ionizing photons from massive short-lived ($<$10 Myr) stars \citep{2022MNRAS.513.2904T}, as it is less affected by dust attenuation compared to rest-frame UV continuum emission, and undergoes a simpler radiation transport compared to the Ly$\alpha$ line.

Consequently, numerous surveys have used the H$\alpha$ luminosity function (LF) to estimate the cosmic star formation rate density (CSFRD) at Cosmic Noon \citep[e.g.,][]{2014ARA&A..52..415M}. To probe fainter galaxy populations, past surveys have generally followed three observational strategies: (1) deep narrow-band imaging with large ground-based telescopes to select H$\alpha$ emitter candidates within narrow specific redshift slices \citep[e.g., ][]{1998ApJ...506..519T, 2012PASP..124..782L, 2013MNRAS.428.1128S}, providing results at discrete redshifts; (2) blind grism spectroscopic surveys using the slitless spectroscopy capability of the Hubble Space Telescope (HST) to identify H$\alpha$ emitters over a broader redshift range \citep[e.g., ][]{2009ApJ...696..785S, 2023ApJ...943....5N}; and (3) spectroscopic follow-up observation after photometric selection in certain band \citep[e.g., ][]{1999MNRAS.306..843G,2007IAUS..235..417R,2013MNRAS.433.2764G}.

A major leap in this work has occurred with
JWST/NIRCam’s unique capabilities for identification of high-redshift galaxies \citep[e.g.,][]{2022ApJ...940L..14N, 2025A&A...696A..87C, 2025arXiv250511263N} and AGNs \citep[including little red dots, LRDs,][]{10.1093/mnras/stad2396, 2025arXiv251103035C, 2024ApJ...963..129M, 2024ApJ...968...38K, 2025arXiv250520393Z}, as well as quantification the LF of H$\alpha$ and other optical emission lines at higher redshifts \citep{2023ApJ...953...53S, 2025ApJ...987..186F}, albeit in a narrower redshift range. At the same time, the JWST/NIRISS grism has also delivered significant improvements in observational depth and wavelength coverage compared to HST/WFC3 \citep[e.g.,][]{2025A&A...699A.225W}, which help us search for fainter H$\alpha$ emitters across a broader redshift range.

Currently, at $z\sim1.3$, existing wide but shallow surveys have mainly constrained the bright end of the H$\alpha$ LF (the characteristic luminosity, $L_{*}$), though the faint-end slope remains poorly determined due to limited depth \citep[e.g.,][]{2013MNRAS.428.1128S,2023ApJ...943....5N}. At $z\sim2.0$, while narrow-band surveys such as \cite{2010A&A...509L...5H} have provided  deep observations, no grism survey to date has measured the H$\alpha$ LF in this redshift range, largely due to the wavelength limitations of the HST/G141 grism. In this work, we present new measurements of the H$\alpha$ LF and CSFRD at $z\sim1.3$ and $z\sim2.0$ using NIRISS Wide-Field Slitless Spectroscopy (WFSS) from the GLASS-JWST Early Release Science Program (ERS-1324, PI: T. Treu) with a new framework estimating survey volume and completeness. Combining JWST’s high sensitivity with the gravitational lensing magnification from the ABELL-2744 cluster, we detect H$\alpha$ emitters down to luminosities of $L_{\rm H\alpha}\sim10^{40.5}\rm~erg~s^{-1}$ at $z\sim1.3$ and $L_{\rm H\alpha}\sim10^{40.9}\rm~erg~s^{-1}$ at $z\sim2.0$, corresponding to SFRs of $\sim$ 0.4 and $\sim$ 1.0 $M_{\odot}\rm~yr^{-1}$, respectively.

This paper is organized as follows: In Section \ref{sec: data}, we describe the observations and data reduction. Section \ref{sec: sample} details the selection of H$\alpha$ emitters and the measurement of their physical properties. Section \ref{sec: LF} presents the processes used to estimate the H$\alpha$ LF and reports the final results. In Section \ref{sec: discussion}, we compare the faint-end slope of LF measurements with previous studies, estimate the CSFRD, and discuss the implications of our methodology. We summarize our conclusions in Section \ref{sec: summary}. Throughout this work, we assume a $\Lambda$CDM cosmology with parameters $\Omega_m$ = 0.30, $\Omega_{\Lambda}$  = 0.7 and $h_0$ = 70 km $\text{s}^{-1}$ $\text{Mpc}^{-1}$, and a \cite{2003PASP..115..763C} initial mass function.

\section{Observations and Data Reduction} \label{sec: data}
In this study, we utilize JWST/NIRISS grism data from the GLASS-ERS program, whose observational strategy is detailed in \cite{2022ApJ...935..110T}. Briefly, the core of the ABELL-2744 cluster (a 130'' $\times$ 130'' region) was observed with NIRISS in WFSS mode for $\sim$18.1 hours, along with $\sim$2.36 hours of direct imaging in three filters (F115W, F150W, and F200W) on 2022 June 28–29 and 2023 July 7. The total exposure times for most sources in the F115W, F150W, and F200W filters are 5.4, 5.7, and 2.9 hours, respectively. These observations provide low-resolution ($R = \lambda/\Delta \lambda~\sim 150$) spectra of all objects in the field of view, covering a wavelength range from 1.0 to 2.2 $\rm \mu m$. This enables the detection of H$\alpha$ emission lines for flux measurements, as well as H$\beta$ and [O~\textsc{III}] lines for redshift determination in the $z$ = 1.0–2.3 range. The grism spectra were obtained at two orthogonal dispersion angles using both the GR150C and GR150R grism elements, reducing contamination from overlapping spectral traces.

For data reduction, we generally follow the GLASS-ERS pipeline described in \cite{2022ApJ...938L..13R}. The entire dataset was processed using the latest set of reference files (\texttt{jwst\_1413.pmap}, which includes recent updates for NIRISS WFSS background corrections) and the Grism Redshift \& Line Analysis software \citep[GRIZLI;][version 1.12]{2021zndo...5012699B}. We began with Stage 0 data (\texttt{uncal.fits}) and used the JWST pipeline (version 1.19) \texttt{Detector1} module to generate count rate maps. We then run the GRIZLI preprocessing pipeline, which performs WCS registration, astrometric alignment, flat-fielding, sky background subtraction, and pixel drizzling, to produce fully reduced individual exposures and mosaics for both the pre-imaging and WFSS datasets. All images were aligned to the LegacySurveys DR10 \citep{2019AJ....157..168D} and Gaia DR3 catalogs \citep{2023A&A...674A...1G}. For image drizzling, we accept mosaics at 30 milli-arcseconds (mas) version for resolved studies of 2D emission line map \citep{2022ApJ...938L..13R}. After that, we performed source detection and automatic photometry using the \texttt{SEP} package on the combined F115W, F150W, and F200W mosaic. The morphological model for each source (which also serves as the reference image for forward modeling) is derived from the segmentation map of the photometry results. Our reduction results are consistent with \cite{2025A&A...699A.225W} data.

We model the contamination for each detected source brighter than $m_{\rm AB}=29$ by fitting a polynomial function to their spectra. All pixels used for extraction and modeling were weighted according to a combination of their contamination level and flux uncertainty, following the setup in \cite{2022ApJ...938L..13R}. To forward-model the 2D grism spectra of galaxies, we adopted Flexible Stellar Population Synthesis (FSPS) templates \citep{2009ApJ...699..486C, 2010ApJ...712..833C}, which include emission line complexes with fixed ratios \citep[see further details in][]{2023ApJS..266...13S}. The 1D spectra were optimally extracted from the 2D data using the morphological model of each source to define the position and spatial extent of the spectral trace.

\section{H$\alpha$ Emitter Samples} \label{sec: sample}
In this section, we construct our H$\alpha$ emitter samples and present their basic properties. Due to the limited spectral resolution of NIRISS, the observed flux is a blend of H$\alpha$ and the [N~\textsc{II}] doublet. Throughout this paper, except in Section 5.2 where we explicitly discuss the H$\alpha$/[N~\textsc{II}] ratio, we follow the convention of \cite{2023ApJ...943....5N} and \cite{2013ApJ...779...34C} and use the terms ``H$\alpha$ sample'', ``H$\alpha$ emission line ($L_{\rm H\alpha}$)'', and ``H$\alpha$ LF'' to denote the blended H$\alpha$+[N~\textsc{II}] measurements.

\subsection{Sample Selection}
\begin{figure*}
\includegraphics[width=1.0\textwidth]{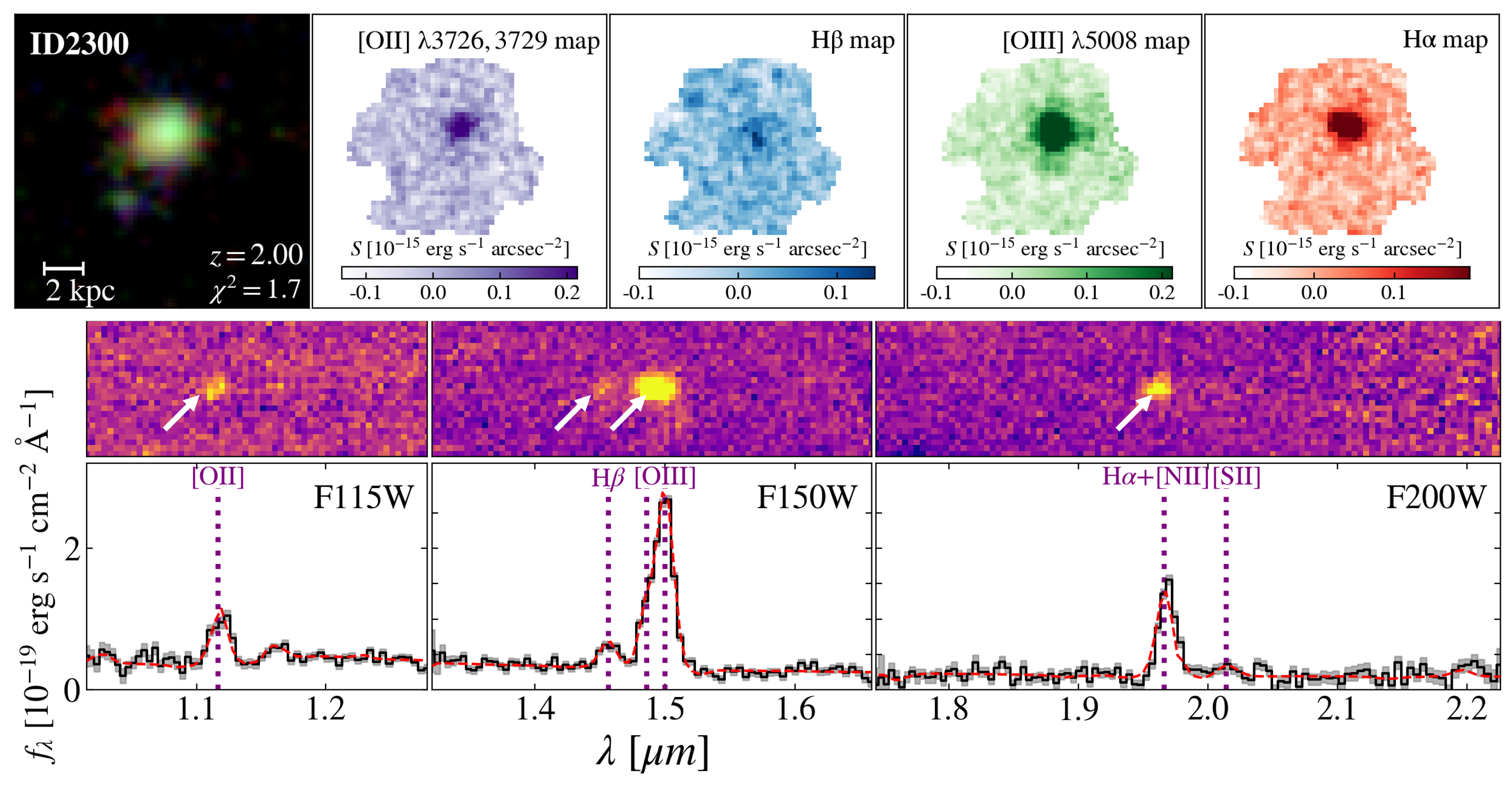}
\caption{ID-2300 as an example to illustrate the 2D emission line maps, 2D spectra, and 1D spectra of the H$\alpha$ emitters in our sample. \textbf{Top panels}, from left to right: a color-composite cutout (RGB image at 30 mas pixel scale; R: F200W, G: F150W, B: F115W), and surface-brightness maps of the emission lines ([O~\textsc{II}], H$\beta$, [O~\textsc{III}], H$\alpha$) derived from the NIRISS slitless spectroscopy. \textbf{Middle panels:} continuum-subtracted 2D grism spectra obtained in the three filters (F115W, F150W, F200W). \textbf{Bottom panels:} The optimally extracted 1D observed flux $F_{\lambda}$ (in units of [$10^{-19}~\rm erg~s^{-1}~cm^{-2}$, its $1\sigma$ uncertainty, and the forward-modeled spectra, shown as the black stepped curves, grey shaded region, and red dash line, respectively. 
The physical scale, redshift, and $\chi^2$ value of the fit are annotated in the upper-left panel.
\label{fig: GRIZLI_fitting}}
\end{figure*}

Our source selection follows methodologies established in previous studies of emission line galaxies from low-resolution grism data \citep{2014ApJ...790..113Z,wangGrismLensAmplifiedSurvey2017,2019ApJ...875..152B,wangCensusSubkiloparsecResolution2020,wangMassMetallicityRelation2022}. To construct our sample of H$\alpha$ emitters, we began by examining the fitted redshift and quality derived from the simultaneous fitting of each galaxy's 2D grism spectrum and broadband photometry using JWST NIRISS observations. We selected galaxies satisfying the following criteria: (1) high enough SNR of the grism spectra, which requires the total infrared magnitude $\rm Mag_{F115W+F150W+F200W} \leq 27$; (2) well-constrained redshift estimates, defined as a 68\% confidence interval $\Delta z = |z_{\rm grism,u68} - z_{\rm grism,l68}| < 0.005 \times (1+z)$; (3) redshift within the range where H$\alpha$ falls within the F115W, F150W, or F200W grism wavelength coverage ($0.54 < z_{\rm grism} < 2.39$); (4) acceptable fitting quality ($\chi^2 < 3$); and (5) robust detection of the H$\alpha$ emission line ($\rm SNR_{H\alpha}>5$). Since objects with secure emission-line detections exhibit significantly smaller redshift uncertainties than our imposed constraint, this criterion effectively eliminates sources with ambiguous continuum-based features or poorly constrained fits from faint objects. Figure \ref{fig: GRIZLI_fitting} presents an example of the fitting results for an H$\alpha$ emitter at $z \sim 2$, where the H$\alpha$ line is detected in the F200W grism, and [O~\textsc{II}] and [O~\textsc{III}] lines are also identified in F115W and F150W, respectively, further confirming the redshift.

From 219 initial candidates, we compiled detailed information for each source. Similarly to \cite{2014ApJ...790..113Z} and \cite{2019ApJ...875..152B}, we include: galaxy’s grism ID number, equatorial coordinates, F115W+F150W+F200W multi-band images, 2D grism spectra at both dispersion angles (in reduced, contamination-subtracted, and continuum-subtracted versions), 1D spectra with best-fit models, probability distributions of redshift, and un-contaminated line maps. We performed careful visual inspection of each candidate, classifying them based on redshift reliability, detection of ancillary emission lines ([O~\textsc{II}], H$\beta$, or [O~\textsc{III}]), and emission-line flux measurement quality.  
Table \ref{tab: emitter} summarizes the results of our examination, listing 99 H$\alpha$ emitters in the ABELL-2744 field with their locations, redshifts, and observed H$\alpha$ fluxes. Most excluded sources suffered from either severe contamination (e.g., proximity to neighboring sources) or showed no clear emission lines in the grism data.

\begin{figure*}
\includegraphics[width=1.0\textwidth]{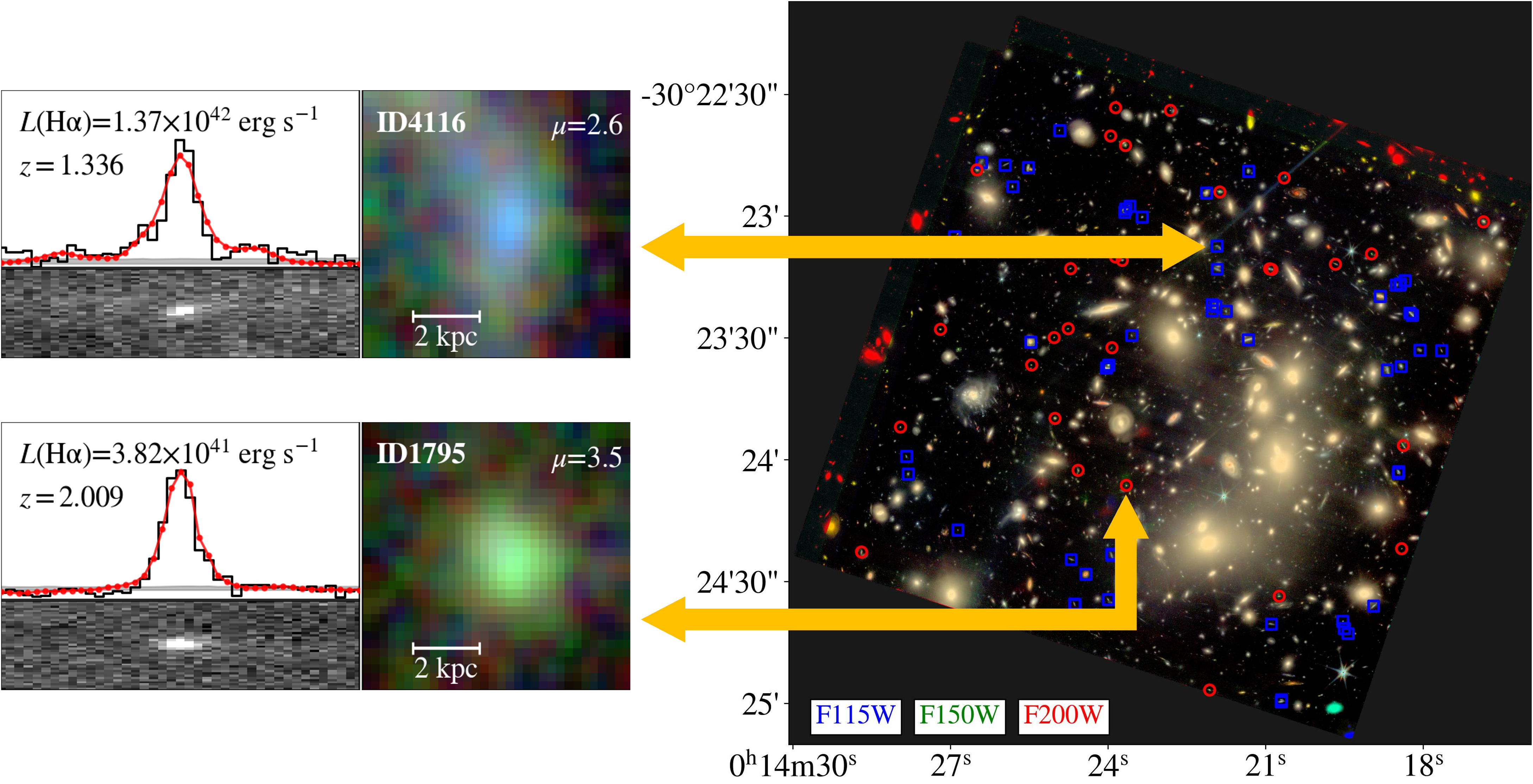}
\caption{H$\alpha$ properties and spatial distribution of our sample. \textbf{Left panel}: Cutout images (right) and H$\alpha$ spectra (left) for two representative galaxies in our sample, ID-4116 and ID-1795. The 1D and 2D spectra are displayed in the upper and lower sub-panels, respectively. The measured H$\alpha$ flux, physical scale, and magnification factor are annotated. \textbf{Right panel:} Spatial distribution of H$\alpha$ emitters at $z\sim1.3$ (blue squares) and $z\sim2.0$ (red circles) in the ABELL-2744 field. The background is the NIRISS RGB image (R: F200W, G: F150W, B: F115W).
\label{fig: emitter}}
\end{figure*}

We recognize the possibility of misidentifying other strong emission lines as H$\alpha$, with the most likely contaminants being higher-redshift [O~\textsc{III}] emitters. 
To quantify the reliability of our sample, we examined the detection rates of additional emission lines within our H$\alpha$ sample. We find that 35 out of 99 emitters show the [O~\textsc{II}] line with $\rm SNR > 3$; 91 out of 99 show the [O~\textsc{III}] line with $\rm SNR > 5$; and 60 sources have a detected [S~\textsc{II}] line with $\rm SNR > 3$. Only 3 sources lack any additional line detection; their redshift reliability is correspondingly lower. Given their small number, however, their inclusion does not significantly bias our final analysis.
To further assess selection completeness, we extracted and examined spectra for a magnitude-limited subsample ($\rm Mag_{F115W+F150W+F200W} \leq 27$) regardless of their initial grism features, confirming no additional high-confidence H$\alpha$ emitters were missed. Figure \ref{fig: emitter} presents two representative galaxies from our H$\alpha$ emitter sample, showing their line profiles, morphologies, and locations within the ABELL-2744 field.

\begin{figure}
\includegraphics[width=0.48\textwidth]{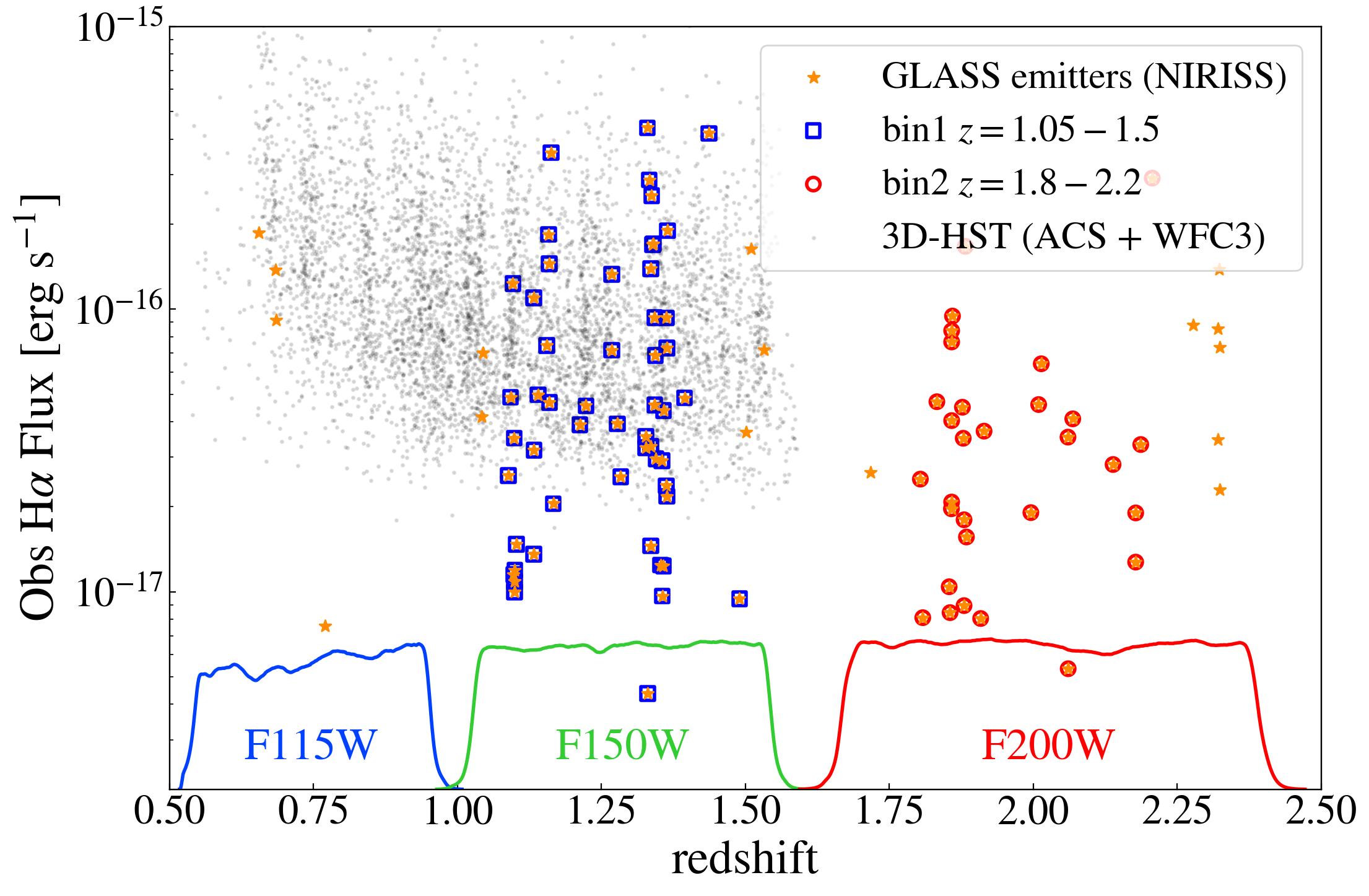}
\caption{Distribution of redshift and observed H$\alpha$ flux for our H$\alpha$ emitter sample (orange stars) at $z=0.6–2.4$. The grey points in the background represent previous H$\alpha$ emitters ($\rm SNR_{H\alpha} > 5$) from the 3D-HST survey \citep{2016ApJS..225...27M}. Thanks to the superior sensitivity of JWST NIRISS, our sample reaches a flux limit 
several times deeper than previous grism surveys 
The colored lines at the bottom show the sensitivity curve of the NIRISS F115W (blue), F150W (green), and F200W (red) filters. The two redshift ranges selected for our luminosity function analysis are highlighted by the blue square ($z=1.05–1.5$) and the red circle ($z=1.8–2.2$).
\label{fig: sample_flux}}
\end{figure}

Figure \ref{fig: sample_flux} compares the redshifts and observed emission-line fluxes of our 99 H$\alpha$ emitters with H$\alpha$ emitters that have the same SNR ($\rm SNR_{H\alpha}>5$) from the 3D-HST data release \citep{2016ApJS..225...27M}, which were observed using the HST G800L and G141 grisms. Thanks to the greater depth and broader wavelength coverage of JWST NIRISS, our sample extends to redshifts beyond 1.8 and reaches a flux limit several times deeper than the 3D-HST survey.

To robustly constrain the luminosity function, we focus on two redshift intervals where the H$\alpha$ emission line falls near the center of the F150W and F200W filters: $z=1.05–1.5$ (53 sources) and $z=1.8–2.2$ (30 sources). This selection, which is based primarily on redshift, is guided by two main considerations.
Firstly, for H$\alpha$ lines detected in the F115W filter, the absence of supporting lines such as [O~\textsc{II}] or [O~\textsc{III}] introduces larger uncertainty in the redshift measurements, which is difficult to quantify in luminosity function estimation. Additionally, the more pronounced spatial extension of lower-redshift galaxies affects both the detection limit and the completeness of H$\alpha$ emitters.
Secondly, we exclude emitters whose H$\alpha$ line falls near the spectral edges of the F150W and F200W filters. Not only does the lower sensitivity reduce the depth at these wavelengths, but part of the spatial extent of the emission line can fall outside the effective spectral trace. This reduces the accuracy of forward modeling and complicates the estimation of effective volume and completeness required for the luminosity function calculation.


\begin{deluxetable}{lcccc}
\tablecaption{\centering H$\alpha$ Emitter Sample in ABELL-2744
\label{tab: emitter}}
\tabletypesize{\footnotesize}
\tablehead{
\colhead{No.} & \colhead{R.A.} & \colhead{Decl.} & \colhead{Redshift}   & \colhead{Observed H$\alpha$ Flux} \\
 & \colhead{deg} & \colhead{deg} & & \colhead{$10^{-18}~\rm erg~s^{-1}~cm^{-2}$} \\
\colhead{(1)} & \colhead{(2)} & \colhead{(3)} & \colhead{(4)} & \colhead{(5)}
}
\startdata
1 & 3.586246 & -30.416542 & 1.268 $\pm$ 0.001 & 71.6 $\pm$ 1.08 \\
2 & 3.58623 & -30.416408 & 1.213 $\pm$ 0.002 & 39.02 $\pm$ 1.31 \\
3 & 3.584109 & -30.416826 & 2.321 $\pm$ 0.001 & 85.12 $\pm$ 1.56 \\
4 & 3.59192 & -30.41576 & 2.060 $\pm$ 0.002 & 35.28 $\pm$ 1.17 \\
5 & 3.58096 & -30.411892 & 1.098 $\pm$ 0.001 & 11.49 $\pm$ 0.54 \\
\enddata
\tablecomments{Column (1): object number. Columns (2)–(3): source coordinates. Column (4): grism redshifts determined by forward modeling. Column (5): observed H$\alpha$ flux with errors (in units of $10^{-18}~\rm erg~s^{-1}~cm^{-2}$), uncorrected for lens magnification.\\
(This table is available in its entirety in machine-readable form in the online article.)}
\end{deluxetable}

\subsection{Lens Models} \label{subsec: lensmodel}

Observations of strongly lensed regions behind galaxy clusters provide access to intrinsically fainter galaxy populations that would otherwise remain undetected in blank-field surveys. However, this advantage comes at the cost of a reduced survey volume and the requirement for precise lens models to derive accurate magnification factors and to identify potential multiple images.

In this work, we adopt the publicly available lens model from \cite{2023ApJ...952...84B}. This model was constructed using the LENSTOOL software \citep{2007NJPh....9..447J} and constrained by new data from GLASS-JWST, UNCOVER, and MUSE observations, which include 149 multiple images ($\sim$66\% more than previous models). The median magnification factors for our sample are 2.5 ($z\sim1.3$) and 2.2 ($z\sim2.0$). 

To assess the impact of different lens models on our results, we compared the fiducial model from \cite{2023ApJ...952...84B} with a comparison model from \cite{2023A&A...670A..60B}. The comparison shows that the survey volume (Section \ref{subsec: Vmax}) in the fiducial model is about 2\% smaller than in the comparison model. The derived intrinsic fluxes show a 4\% scatter between the two models, leading to negligible differences in the estimated completeness and flux bias corrections (Section \ref{subsec: completeness}). Consequently, the faint-end slope of the luminosity function (Section \ref{subsec: LF}) differs by about 0.01 between the two models, and the cosmic star formation rate density (Section \ref{subsec: CSFRD}) derived from the fiducial model is about 1\% higher than that from the comparison model. These systematic differences are negligible compared to the statistical uncertainties presented in the following sections. We therefore conclude that the choice of lensing model does not lead to significantly different conclusions.

\section{H$\alpha$ Luminosity Functions} \label{sec: LF}
We derive the H$\alpha$ LFs using the direct $1/V_{\max}$ method
\citep{1968ApJ...151..393S}, expressed as

\begin{equation}
\Phi(L)=\frac{1}{d~\log L} \sum_{i} \frac{1}{C_{i}V_{\max,i}}
\end{equation}

\begin{figure*}
\includegraphics[width=1.0\textwidth]{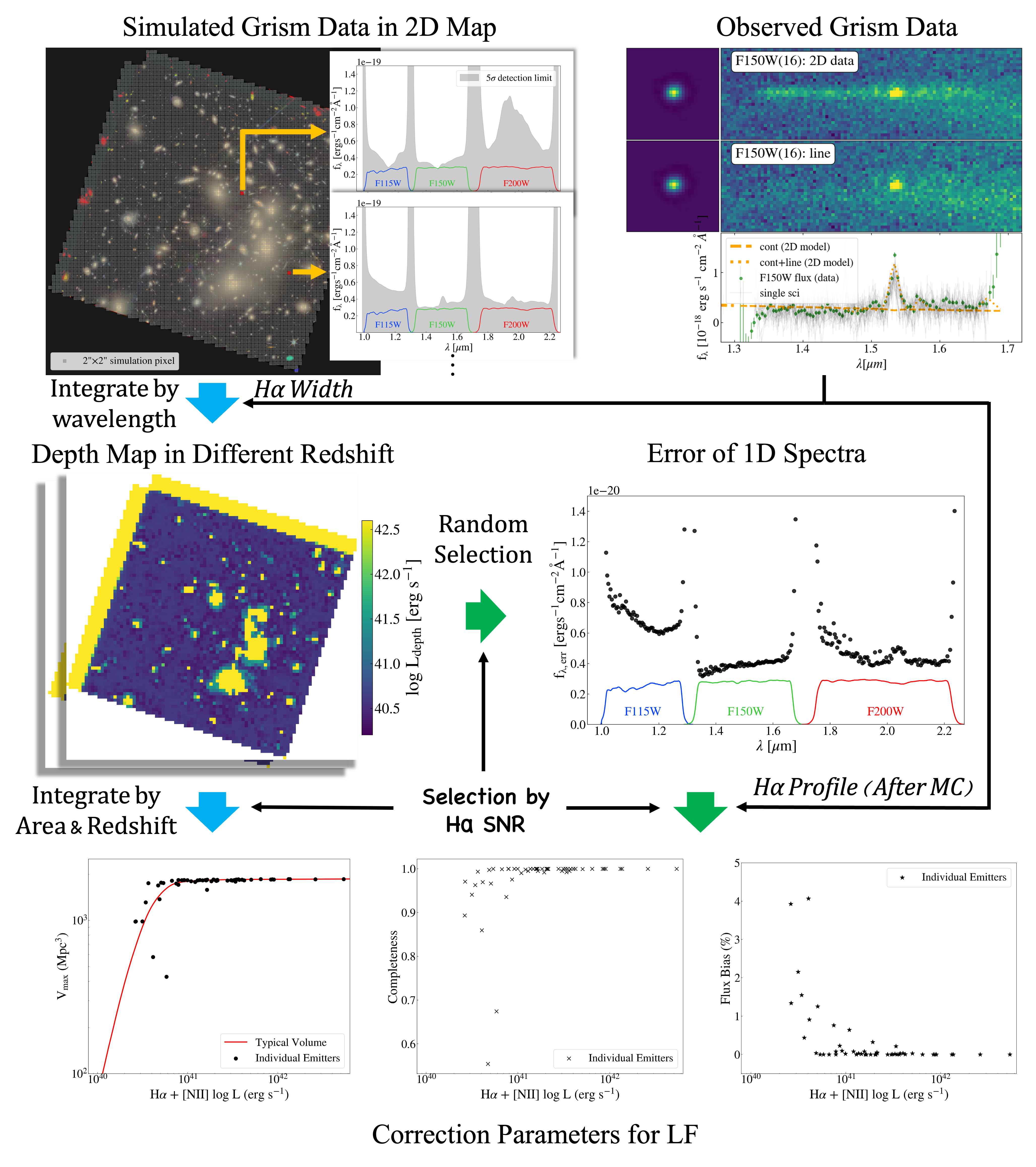}
\caption{Estimation of the correction parameters for the luminosity function calculation. The blue and green arrows are the main steps for survey volume (Section \ref{subsec: Vmax}) and completeness \& flux bias (Section \ref{subsec: completeness}), respectively, while the black arrows represent the auxiliary data or criteria applied within those steps. \textbf{Top panels: Input data.} Left: Schematic of the segmentation and simulated extraction performed for each 2"$\times$2" position in the ABELL-2744 field to get the extracted spectra as well as their error in each location. Right: An example of an actual 2D grism data and its corresponding 1D spectrum around the H$\alpha$ emission line. \textbf{Middle panels: Depth and noise analysis.} Left: The 2D depth map derived from the H$\alpha$ with emission line width of 116$\rm \AA$ in the observed spectra. Right: The flux error distribution of the 1D spectra randomly selected in pixels that satisfied the SNR selection.
\textbf{Bottom panels: Derived parameters.} Left: The effective survey volume, computed by integrating the area which satisfied the SNR selection in each depth map and luminosity distance over the relevant redshift range. Middle and Right: The completeness and flux bias value, derived by selection results from the mock spectra that injected the H$\alpha$ emission lines after Monte Carlo simulations to the extracted spectra.
\label{fig: simulation_logic}}
\end{figure*}

Before performing the final LF calculation and model fitting, the relevant parameters for each H$\alpha$ emitter (including $V_{\max,i}$, the maximum observable volume for the $i$-th source; and $C_i$, the completeness of the $i$-th source in the luminosity bin) need to be determined. In this section, we first describe in detail the procedure for estimating these parameters for each source (as illustrated in Figure \ref{fig: simulation_logic}), and then present the corrected LF results (as shown in Figure \ref{fig: LF_results}).

\subsection{Survey Volume} \label{subsec: Vmax}
Due to the moderate sample size of our Ha sample, we estimate the survey volume and completeness for each source. We begin with the simulated error cube for each position \& wavelength within the GLASS-JWST field of view. Across the entire survey footprint, we apply the same extraction procedure implemented by GRIZLI for actual emitters, obtaining 1D grism spectra in the F115W, F150W, and F200W bands over a 2.0''$\times$2.0'' grid (as shown in the upper-left panel in Figure \ref{fig: simulation_logic}). During extraction, contamination models are incorporated into each 1D spectrum's error using the same weighting formula applied in the GRIZLI fitting process \citep{2022ApJ...938L..13R}. As a result, regions affected by bright contamination exhibit correspondingly larger uncertainties in their 1D spectra at the relevant wavelengths.

We measure the spectral full width at half maximum (FWHM) of each H$\alpha$ emission line from its observed 1D spectrum. The detection limit at each wavelength is then computed by co-adding the 1D errors within a window of $\pm$FWHM around the line center. Since most H$\alpha$ emission lines at cosmic noon are spatially resolved, their line widths, reflecting the physical extent of the galaxies, show considerable variation. Adopting a uniform median FWHM value for each redshift bin would therefore introduce significant bias in the volume estimation, as illustrated by the discrepancy between the individual measurements (black points) and the median value (red line) in the lower-left panel of Figure \ref{fig: simulation_logic}.

Through the above procedure, we construct a magnification-corrected 5$\sigma$ depth map at a resolution of 2.0''$\times$2.0'' for each H$\alpha$ width in the redshift intervals $z=1.05–1.5$ and $z=1.8–2.2$, using a step size of $\Delta z=0.01$ (see the left middle panel of Figure \ref{fig: simulation_logic} for an example). For each source with an observed H$\alpha$ luminosity, we identify the region in the image plane where its magnified flux would exceed the local 5$\sigma$ detection limit at each redshift step. The corresponding survey area in the source plane is derived by dividing the image-plane area by the magnification factor at each position. The total survey volume is finally obtained by integrating the source-plane area over each redshift bin.

The relationship between survey volume and intrinsic (magnification-corrected) H$\alpha$ luminosity for sources at $z\sim1.3$ is shown in the lower-left panel of Figure \ref{fig: simulation_logic}. In general, the survey volume increases with intrinsic H$\alpha$ luminosity, flattening at the luminosity limit corresponding to a magnification factor of $\mu=1$.

\subsection{Sample Completeness \& Flux Bias} \label{subsec: completeness}
A robust determination of the luminosity function requires a careful assessment of the detection completeness. Previous studies have demonstrated that GRIZLI effectively identifies emission line components through its forward modeling method \citep{wangDiscoveryStronglyInverted2019,2021ApJ...920...78B}. Meanwhile, noise fluctuations across the survey area are already incorporated in the volume calculation described in Section \ref{subsec: Vmax}. Regarding the redshift selection completeness, our H$\alpha$ emission lines fall within either the F150W ($z\sim1.3$) or F200W ($z\sim2.0$) filters, enabling the grism spectra from F115W or F150W to detect additional lines (such as [O~\textsc{II}], H$\beta$, and [O~\textsc{III}]) for redshift verification. For further verification, we also extract and examine spectra for a magnitude-limited subsample ($\rm Mag_{F115W+F150W+F200W}\leq27$), regardless of their initial grism features, confirming the absence of any additional high-confidence H$\alpha$ emitters.

We evaluate the H$\alpha$ line detection completeness in grism spectra using a Monte Carlo (MC) approach. 
First, we randomly select 200 locations from the regions in the image plane where the magnified flux of a given source exceeds the 5$\sigma$ detection limit from the depth map at its redshift, and extract the corresponding 1D spectra along with spectral errors.
At each location, we then inject 100 mock emission lines into the extracted 1D spectra (as illustrated in the middle-right panel of Figure \ref{fig: simulation_logic}), which are generated by the following processes:
\begin{itemize}
    \item [1.] The observed, magnification-corrected flux and the corresponding 1D spectrum (with error) of the H$\alpha$ emission line are collected for each source.
    \item [2.] Derived the simulated total flux from a Gaussian distribution centered on the observed flux, with a standard deviation given by its flux error.
    \item [3.] Scaled the observed 1D H$\alpha$ spectral profile and its error by the ratio of simulated total flux to the original observed flux.
    \item [4.] Generated the mock 1D emission line profile by sampling from a Gaussian distribution centered on the scaled profile, with a standard deviation given by its error at each wavelength.
    \item [5.] Applied the lensing magnification in each location to the mock profile.
\end{itemize}

From the 20,000 simulated spectra (200 locations $\times$ 100 mock profiles), we compute the fraction of 5$\sigma$ detections, which defines the completeness for a given emitter. The relationship between completeness and intrinsic (magnification-corrected) H$\alpha$ luminosity for sources at $z\sim1.3$ is shown in the bottom-middle panel of Figure \ref{fig: simulation_logic}. Although some scatter is present due to variations in line profiles and local contamination, the completeness increases with intrinsic H$\alpha$ luminosity as expected.

The flux measurements of low-SNR emission lines can be artificially boosted due to random noise fluctuations, an effect commonly referred to as the Eddington bias \citep{1913MNRAS..73..359E}. In this work, we follow the methodology established in recent H$\alpha$ LF studies \citep{2023ApJ...953...53S, 2025ApJ...987..186F}, using the observational data to quantify this effect. Specifically, we first build up the mock spectra for each source using the same procedure for the completeness estimation (200 locations $\times$ 100 mock profiles). Then we apply the 5$\sigma$ H$\alpha$ detection criterion to these mock spectra. Due to random flux fluctuations, some mock H$\alpha$ lines fall below the detection threshold, causing the median flux of the detected mocks to be higher than the input value. The median flux ratio between all input mock lines and those recovered was adopted as the flux correction factor for the simulated source.
As shown in the lower right panel in Figure \ref{fig: simulation_logic} ($z\sim1.3$), the derived correction factor for the Eddington bias is found to be smaller than 5\% even at the faint end of our luminosity range. Consequently, our sample is less affected by the Eddington bias compared to high-redshift H$\alpha$ emitter samples \citep{2023ApJ...953...53S, 2025ApJ...987..186F}. This can be attributed to the fact that the H$\alpha$ emission lines in our sample are generally more spatially extended, and the lower resolution of NIRISS grism data relative to NIRCam WFSS that necessitates a higher SNR threshold for secure H$\alpha$ emitter detection.

\subsection{Multiply Imaged Systems}
Using the strong gravitational lensing model described in Section \ref{subsec: lensmodel}, we computed and performed positional matching of multiple images for all selected sources. The results indicate that most of our selected H$\alpha$ emitters have only a single lensed image within the GLASS-JWST NIRISS field of view. The only exception is a triply-imaged galaxy at $z\sim2.01$, which was previously reported by \cite{2025A&A...699A.225W}. Our sample contains two of its three images (IDs 615 and 1795, corresponding to $\rm ID_{Watson}$ 380 and 998, respectively). In the final LF estimation, these two images are treated as a single source, and the intrinsic flux is estimated as the average of the two images' intrinsic fluxes, weighted by inverse variance. Furthermore, we visually inspected images of all H$\alpha$ emitters and confirmed that none of them had been split into separate sources due to shear effects in the strong gravitational field. 

\subsection{LF results} \label{subsec: LF}

\begin{deluxetable}{ccccc}
\tablecaption{\centering H$\alpha$ Luminosity Functions
\label{tab: LF}}
\tabletypesize{\footnotesize}
\tablehead{
\colhead{$\log L_{\rm H\alpha}$} & \colhead{N} & \colhead{$<\rm c>$} & \colhead{$<V_{\max}>$}   & \colhead{$\log \Phi_{\rm H\alpha}$} \\
\colhead{($\rm erg~s^{-1}$)} &  &  & \colhead{($/V_{\max}(L_{\max})$)} & \colhead{($\rm Mpc^{-3}~dex^{-1}$)} \\
\colhead{(1)} & \colhead{(2)} & \colhead{(3)} & \colhead{(4)} & \colhead{(5)}
}
\startdata
\multicolumn{5}{c}{$z\sim1.3$} \\
\hline
$<$40.5 & 3 & 0.90 & 0.46 &  \\
40.75 $\pm$ 0.25 & 16 & 0.93 & 0.78 & $-1.57^{+0.13}_{-0.14}$ \\
41.25 $\pm$ 0.25 & 18 & 0.99 & 0.97 & $-2.00^{+0.12}_{-0.14}$ \\
41.75 $\pm$ 0.25 & 12 & 1.00 & 0.99 & $-2.19^{+0.15}_{-0.17}$ \\
42.25 $\pm$ 0.25 &  3 & 1.00 & 1.00 & $-2.79^{+0.30}_{-0.38}$ \\
42.75 $\pm$ 0.25 &  1 & 1.00 & 1.00 & $-3.27^{+0.52}_{-1.06}$ \\
\hline
\multicolumn{5}{c}{$z\sim2.0$} \\
\hline
$<$40.9 & 1 & 0.34 & $<$0.20 &  \\
41.10 $\pm$ 0.20 & 8 & 0.83 & 0.51 & $-1.92^{+0.21}_{-0.24}$ \\
41.50 $\pm$ 0.20 & 13 & 0.92 & 0.95 & $-2.06^{+0.13}_{-0.15}$ \\
41.90 $\pm$ 0.20 &  5 & 0.98 & 0.97 & $-2.44^{+0.23}_{-0.28}$ \\
42.30 $\pm$ 0.20 &  2 & 1.00 & 1.00 & $-2.90^{+0.37}_{-0.56}$ \\
\enddata
\tablecomments{Column (1): logarithm of each H$\alpha$ luminosity bin. (2): number of H$\alpha$ emitters in each bin. (3): mean completeness of H$\alpha$ emitters in each bin. (4): mean volume in which the H$\alpha$ emitter with the given intrinsic luminosity can be detected, compared with the total survey volume in each bin. (5): number density, with the uncertainty including both small-number Poisson error and cosmic variance.}
\end{deluxetable}

The observed H$\alpha$ LFs are estimated at $z\sim1.3$ and $z\sim2.0$. The final sample used for the LF measurement consists of 53 and 29 individual galaxies at $z\sim1.3$ and $z\sim2.0$, respectively, giving a total of 82 sources.

\begin{figure*}
\includegraphics[width=1.0\textwidth]{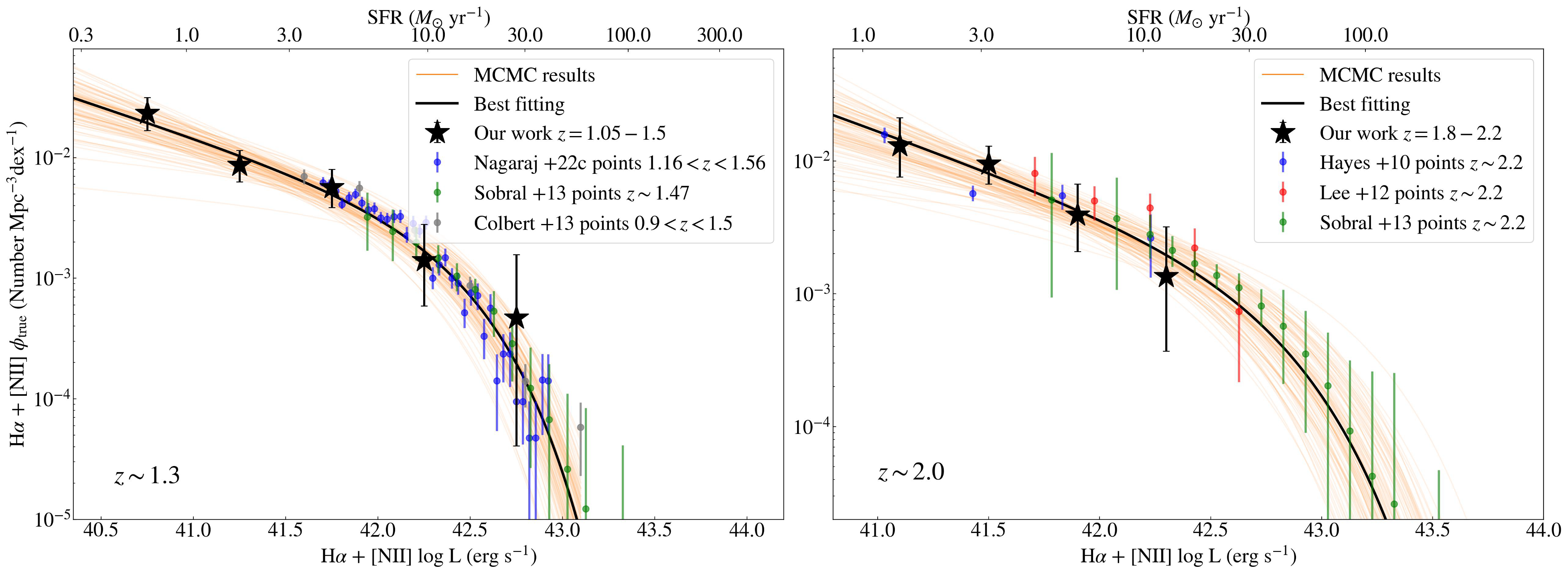}
\caption{Observed H$\alpha$ luminosity functions at $z\sim1.3$ (left) and 
$z\sim2.0$ (right). The black solid line represents the best-fit Schechter function, while the orange lines in the background show 100 randomly selected samples from the MCMC chains. The fitting procedure incorporates H$\alpha$ LF results from \cite{2023ApJ...943....5N} at $z\sim1.3$ and \cite{2013MNRAS.428.1128S} $z\sim2.0$ as priors. The top axis indicates the corresponding SFR from H$\alpha$ flux \citep{2012ARA&A..50..531K} based on the [N~\textsc{II}]/H$\alpha$ flux ratio from \cite{2018ApJ...855..132F} and after extinction correction. Literature results from narrow-band \citep{2010A&A...509L...5H,2012PASP..124..782L} and WISP grism surveys \citep{2013ApJ...779...34C} are included for comparison. All results are scaled to a \cite{2003PASP..115..763C} IMF and dust corrected assuming $A_{\rm H\alpha}=1$.
\label{fig: LF_results}}
\end{figure*}

The derived LFs are listed in Table \ref{tab: LF} and shown as black stars in Figure \ref{fig: LF_results} after completeness and flux bias correction. When calculating the errors, we consider Poisson noise for the small number statistics \citep{1986ApJ...303..336G} and also include the uncertainty caused by cosmic variance following the methods in \cite{2011ApJ...731..113M}, which estimates large-scale density fluctuations as the product of the galaxy bias \citep{2018MNRAS.481.4885Q,2024MNRAS.533.2391D}.

We fit the H$\alpha$ LFs using a Schechter function defined as:

\begin{equation}
\Phi(L)=\ln(10)\phi^*(\frac{L}{L^*})^{\alpha+1}e^{-L/L^*}
\end{equation}

where $\phi^*$ represents the number density normalization, $L^*$ is the characteristic (break) luminosity, and $\alpha$ is the faint-end slope.

The Schechter function is fitted using a Monte Carlo Markov Chain (MCMC) approach implemented with the \texttt{emcee} package for the binned LF results shown in Table \ref{tab: LF}. We find that the characteristic luminosity $L^*$ is poorly constrained due to the limited survey volume and the small number of bright sources. There is also strong degeneracy between $L^*$ and the faint-end slope $\alpha$ if all three Schechter parameters are left free. To address this, we adopt reasonable parameter ranges ($40<\log L<45$, $-3<\alpha<-1$, and $-5<\log \phi^*<-1$, and incorporate Gaussian distributions priors, with the mean and sigma set to the best-fit values and uncertainties from previous results (\citealt{2023ApJ...943....5N} at $z\sim1.3$ and \citealt{2013MNRAS.428.1128S} $z\sim2.0$) in the MCMC sampling.

The best-fit Schechter functions, along with a subset of MCMC realizations, are shown in Figure \ref{fig: LF_results}. The corresponding parameter values are summarized in Table \ref{tab: LF_para}. It is worth noticing that since the Schechter function parameters are not independent, assuming Gaussian distribution priors may slightly overestimate the error bars, but have a negligible effect on the parameter values. Further discussions are presented in the next section.

\begin{deluxetable}{cccc}
\tablewidth{300\%}
\tablecaption{\centering Schechter Fit Parameters of the H$\alpha$ Luminosity Functions
\label{tab: LF_para}}
\tabletypesize{\footnotesize}
\tablehead{
\colhead{$z$} & \colhead{$\log L^{*}_{\rm H\alpha}$} & \colhead{$\log \Phi^{*}_{\rm H\alpha}$} & \colhead{$\alpha_{\rm H\alpha}$} \\
  & ($\rm erg~s^{-1}$) & \colhead{($\rm Mpc^{-3}~dex^{-1}$)} &  \\
\colhead{(1)} & \colhead{(2)} & \colhead{(3)} & \colhead{(4)}
}
\startdata
1.3 & $42.39^{+0.06}_{-0.08}$ & $-2.89^{+0.13}_{-0.09}$ & $-1.50^{+0.14}_{-0.08}$ \\
2.0 & $42.73^{+0.11}_{-0.13}$ & $-3.17^{+0.14}_{-0.09}$ & $-1.60^{+0.17}_{-0.09}$ \\
\enddata
\end{deluxetable}

\section{Discussion} \label{sec: discussion}
In this section, we compare our estimates of Cosmic SFR Density and faint-end slope $\alpha_{\rm H\alpha}$ with previous work, analyze the potential AGN contamination, and discuss the implications of our work.

\subsection{Faint-end Slope of LF} \label{subsec: Faint-end}
\begin{figure*}
\includegraphics[width=1.0\textwidth]{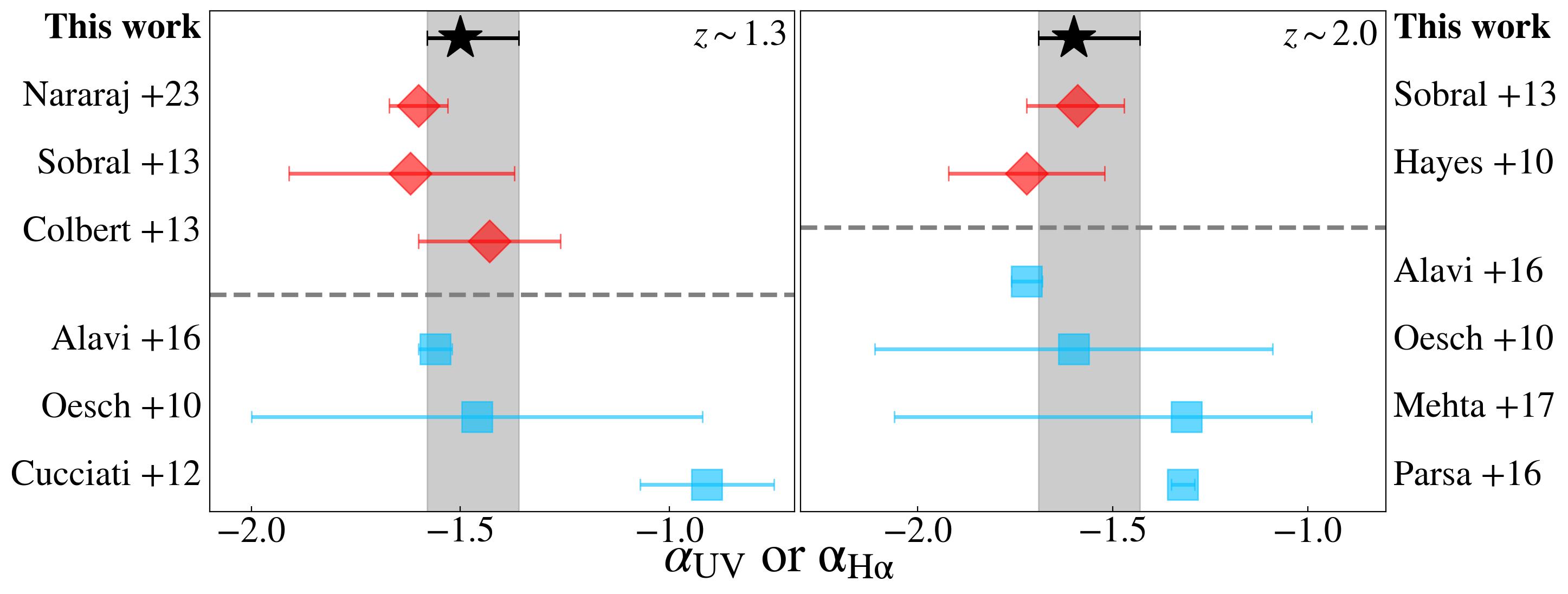}
\caption{The 16\%–50\%–84\% credible interval constraints on the faint end slope of UV and H$\alpha$ LF in $z\sim1.3$ (left) and $z\sim2.0$ (right). Our results (black stars) are compared with previous results for the H$\alpha$ LF (red points) \citep{2010A&A...509L...5H, 2013ApJ...779...34C, 2013MNRAS.428.1128S, 2023ApJ...943....5N} and the UV LF (blue points) \citep{2010ApJ...725L.150O, C2012, 2016MNRAS.456.3194P, 2016ApJ...832...56A, 2017ApJ...838...29M}.
\label{fig: faint_end_slope}}
\end{figure*}

In this work, we have probed the H$\alpha$ luminosity function down to approximately $\sim 0.01L_*$ using grism data for the first time, enabling robust constraints on the faint-end slope ($\alpha_{\rm H\alpha}$). Figure \ref{fig: faint_end_slope} compares our derived $\alpha_{\rm H\alpha}$ with previous determinations from H$\alpha$ LFs (based on narrowband imaging or HST grism spectroscopy) and UV LFs.

For both redshift bins, our measured $\alpha_{\rm H\alpha}$ is consistent with previous H$\alpha$ LF estimates. 
Specifically, at $z\sim1.3$, our LF data points in the luminosity range $10^{41.5}$–$10^{43}$ erg s$^{-1}$ agree well with literature results. Our faint-end measurements also provide crucial support for previous determinations of the slope, which range from –1.43 to –1.62 \citep{2013ApJ...779...34C, 2013MNRAS.428.1128S, 2023ApJ...943....5N}. Considering that the [O~\textsc{III}]/H$\alpha$ flux ratio tends to increase toward lower-luminosity galaxies \citep[e.g.,][]{2013ApJ...779...34C,2019MNRAS.490.3667Z}, the faint-end slope of the [O~\textsc{III}] LF is expected to be steeper than the H$\alpha$ LF. Our measured H$\alpha$ slope of –1.50 is therefore consistent with the value of $\alpha_{\rm [O,III]} =-1.6$ that has often been assumed at similar redshifts \citep[e.g.,][]{2015MNRAS.451.2303S}.
At $z\sim2.0$, our results are in good agreement with the very deep narrowband photometric study of \cite{2010A&A...509L...5H}, confirming the consistency between selection methods using slitless spectroscopy and narrow-band imaging.

We measured the faint-end slopes of $-1.50^{+0.14}_{-0.08}$ at $z\sim1.3$ and $-1.60^{+0.17}_{-0.09}$ at $z\sim2.0$. Comparing this with recent high-redshift studies \citep[e.g.,][]{2023ApJ...946..117B, 2025ApJ...987..186F}, which report slopes around –1.7 at z=4.5–6.2, suggests at most weak evolution in $\alpha_{\rm H\alpha}$ 
from cosmic noon to the epoch of reionization. It is worth noting that the characteristic luminosity $L_{\rm H\alpha}^*$ exhibits a more pronounced positive correlation with redshift.

The faint-end slope comparison between H$\alpha$ and UV LFs is more complicated. Theoretically, multiple competing effects influence the faint-end slope comparison. One of them is the dust attenuation. Recent ALMA observations have confirmed that dust content correlates with stellar mass and SFR \citep{2020A&A...644A.144D}. For standard extinction curves \citep[e.g.,][]{1999PASP..111...63F, 2000ApJ...533..682C}, rest-frame UV light experiences stronger attenuation than H$\alpha$ emission. Consequently, if a uniform dust correction is applied, the UV LF is expected to be steeper than the H$\alpha$ LF. On the other hand, galaxies with lower UV luminosities and lower SFR tend to exhibit higher H$\alpha$-to-UV luminosity ratios \citep[$L_{\rm H\alpha}/L_{\rm UV}$
; e.g.,][]{2017MNRAS.465.3637M, 2020MNRAS.493.5120M, 2022MNRAS.511.4464A}. This is likely due to (i) a higher ionizing photon production efficiency $\xi_{\rm ion}$ at lower metallicities \citep[e.g.,][]{2019MNRAS.486.2215K}, and (ii) more bursty star formation histories in low-mass galaxies \citep[e.g.,][]{2014ApJ...792...99S, 2019ApJ...881...71E}. These effects would lead to a steeper faint-end slope for the H$\alpha$ LF compared to the UV LF. Other additional factors, including IMF \citep{2008ApJ...675..163H}, may also influence the LF results.

In practice, the interplay of these factors makes direct theoretical predictions challenging. As shown in Figure \ref{fig: faint_end_slope}, our measured faint-end slope is broadly consistent with most previous UV LF studies. The flatter slopes reported in some UV LF works are likely due to insufficient depth in UV detections, leading to degeneracy in the fitting parameters. To fully resolve this issue, constructing a UV-complete sample is essential for an unbiased comparison of physical properties with the H$\alpha$-selected sample.

\subsection{Estimation of Cosmic SFR Density} \label{subsec: CSFRD}
To derive the SFR from the H$\alpha$ + [N~\textsc{II}] emission, we applied a set of relations simultaneously. First, the SFR was derived from the dust-corrected H$\alpha$ luminosity using the relation from \cite{2012ARA&A..50..531K}:
\begin{equation} \label{eq: SFR}
\begin{split}
\log [{\rm SFR}/(M_{\odot}~{\rm yr}^{-1})]=& \log[L_{\rm H\alpha}/({\rm erg~s^{-1}})]-41.27 \\
&+\log(0.94),
\end{split}
\end{equation}
suitable for the 
\cite{2003PASP..115..763C} initial mass function \citep[IMF,][]{2014ARA&A..52..415M}. To account for the contribution of [N~\textsc{II}] lines, we adopted the empirical relation among [N~\textsc{II}]/H$\alpha$ flux ratio, stellar mass ($M_*$), and redshift from \cite{2018ApJ...855..132F}:
\begin{equation}
\begin{split}
\log [M_{*}/M_{\odot}] =& 3.696 \xi + 3.236\xi^{-1}+0.729\xi^{-2}\\
&+14.928+0.156(1+z)^2
\end{split}
\end{equation}
where $\xi=\log([\mathrm{N~II}]/\mathrm{H}\alpha)+0.138-0.042(1+z)^2$. Since the ratio depends on $M_*$, we assume that galaxies follow the star-forming main sequence relation from \cite{2014ApJS..214...15S} to connect SFR to $M_*$:
\begin{equation}
\begin{split}
\log [{\rm SFR}/(M_{\odot}~{\rm yr}^{-1})] =& (0.84-0.026t)\log [M_{*}/M_{\odot}]-\\
&(6.51-0.11t)
\end{split}
\end{equation}
where $t$ is the age of the universe in Gyr. For short, the three unknown parameters are the [N~\textsc{II}]/H$\alpha$ flux ratio, the SFR, and the stellar mass, which are constrained by the three functions upon.
Finally, the dust-corrected H$\alpha$ luminosity is related to the observed H$\alpha$ luminosity by:
\begin{equation} \label{eq: AHa}
\begin{split}
\log[L_{\rm H\alpha}/({\rm erg~s^{-1}})] = \log[L_{\rm H\alpha,~obs}/({\rm erg~s^{-1}})] + 0.4A_{\rm H\alpha}
\end{split}
\end{equation}
where $A_{\rm H\alpha}$ refer to the H$\alpha$ extinction value. To estimate that value, we select 32 sources in our sample with $\rm SNR_{H\beta}>3$ and without severe blending with the [O~\textsc{III}] $\lambda 4959$ line. Assuming Case B recombination with an electron temperature $T_e = 10^4~\rm K$ and density $n_e = 100~\rm cm^{-3}$, the intrinsic (unattenuated) H$\alpha$/H$\beta$ ratio is 2.86 \citep[e.g.,][]{2006agna.book.....O}. We then use such ratio to constraint the $A_{\rm H\alpha}$ parameter:
\begin{equation}
\begin{split}
\log[L_{\rm H\alpha}/({\rm erg~s^{-1}})] =  \log(2.86) + \log[L_{\rm H\beta}/({\rm erg~s^{-1}})]
\end{split}
\end{equation}
\begin{equation}
\begin{split}
\log[L_{\rm H\beta}/({\rm erg~s^{-1}})] = \log[L_{\rm H\beta,~obs}/({\rm erg~s^{-1}})] + 0.4A_{\rm H\beta}
\end{split}
\end{equation}
where $A_{\rm H\beta}$ is related to $A_{\rm H\alpha}$ by the extinction curve from \cite{1999PASP..111...63F}. We derive $A_{\rm H\alpha} = 0.94^{+0.14}_{-0.22}$ (all 32 sources locate in the 3$\sigma$ range) from the mean observed H$\alpha$/H$\beta$ ratio of 4.06. This is consistent with the spatially resolved analysis of emission-line galaxies in the ABELL-2744 field \citep{Ren.2026}, and previous estimations in a similar redshift \citep{2010A&A...509L...5H, 2013MNRAS.428.1128S}, which also obtain a similar value of $A_{\rm H\alpha} \sim 1$ from the H$\alpha$/H$\beta$ map.  Thus, we adopted $A_{\rm H\alpha} = 1.0$ mag for the H$\alpha$ emission line for all samples in Equation \ref{eq: AHa}. After that, the relations \ref{eq: SFR} to \ref{eq: AHa} provide the conversion from H$\alpha$ + [N~\textsc{II}] flux to SFR, which we applied to our H$\alpha$ luminosity function (see the upper axis in Figure \ref{fig: LF_results} for the corresponding SFR scale).

We computed the cosmic star formation rate density (CSFRD, $\rho_{\rm SFR}$) by integrating the star formation rate function (SFRF), which was derived from the best-fit Schechter function of the H$\alpha$ LF. The lower integration limit was set at $0.4 M_{\odot}~\rm yr^{-1}$ ($L_{\rm H\alpha}=10^{40.5}~\rm erg~s^{-1}$), corresponding to an absolute UV magnitude limit of $M_{\rm UV}=-17.6$ mag after attenuation. For each redshift bin, we performed MC sampling of the posterior distributions of the SFRF parameters and integrated to determine the 16th, 50th, and 84th percentiles of ($\rho_{\rm SFR}$). Our results show that ($\rho_{\rm SFR}$) increase from $0.097^{+0.015}_{-0.016}~M_{\odot}~\rm yr^{-1}~Mpc^{-3}$ at $z\sim1.3$ to $0.129^{+0.025}_{-0.030}~M_{\odot}~\rm yr^{-1}~Mpc^{-3}$ at $z\sim2.0$, confirming the peak of cosmic star formation activity around $z\sim2.0$.

\begin{figure*}
\includegraphics[width=1.0\textwidth]{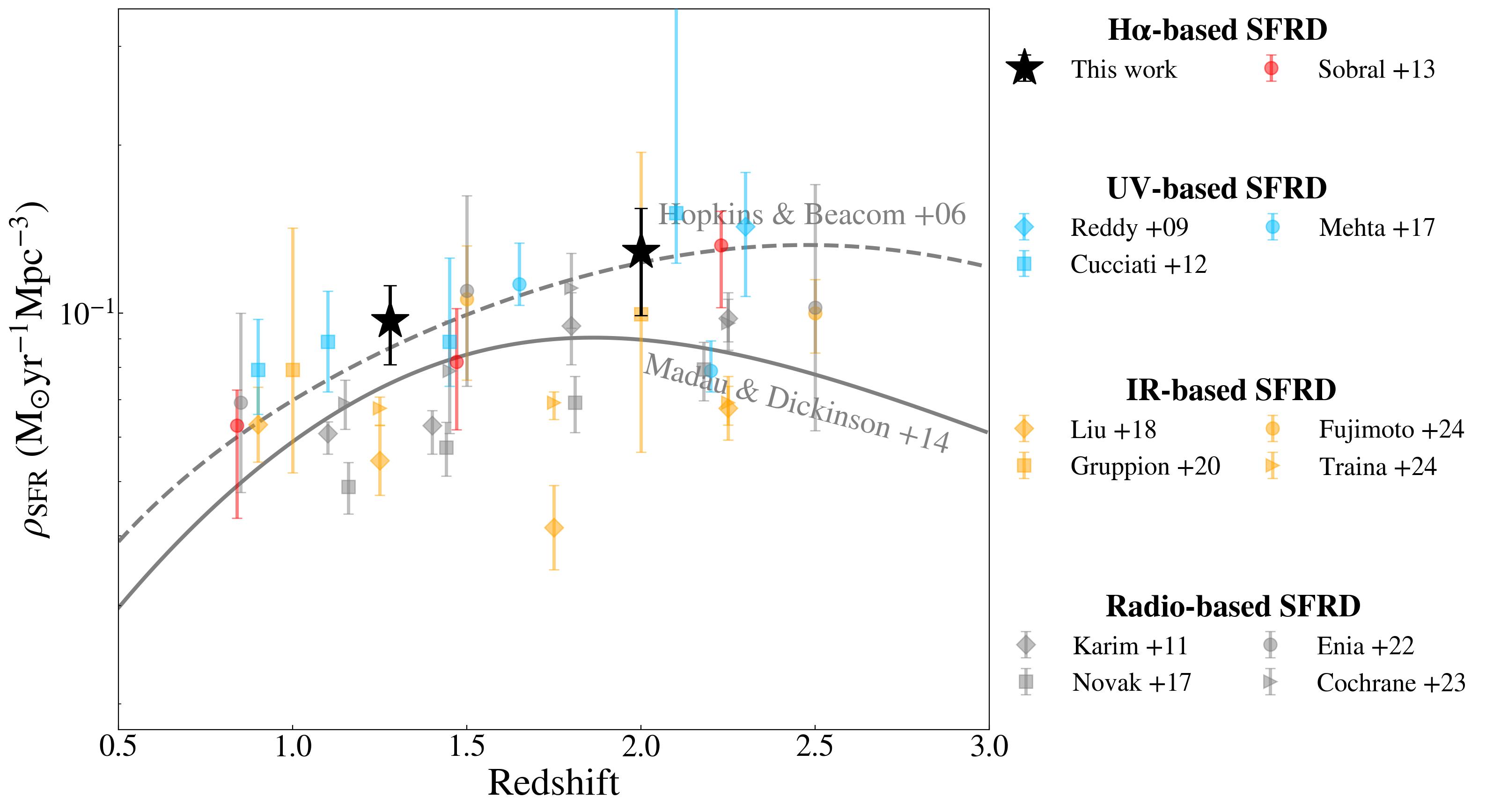}
\caption{Estimates of cosmic SFR density at cosmic noon. Our results are compared to previous measurements based on: H$\alpha$ surveys \citep{2013MNRAS.428.1128S}; dust-corrected UV LFs \citep{2009ApJ...692..778R, C2012, 2017ApJ...838...29M}; infrared surveys \citep{2018ApJ...853..172L, G2020, 2024ApJS..275...36F, T2024}; and radio surveys \citep{2011ApJ...730...61K, N2017, 2022ApJ...927..204E, 2023MNRAS.523.6082C}. The solid and dashed gray lines are the fits by \cite{2014ARA&A..52..415M} and \cite{2006ApJ...651..142H}, respectively. All literature results were corrected for a \cite{2003PASP..115..763C} IMF, the lower limits of SFR adopted for the CSFRD integration are listed in Table \ref{tab: CSFRD_lower_limit}.
\label{fig: SFRD}}
\end{figure*}

Figure \ref{fig: SFRD} compares our CSFRD measurements with previous works. To ensure consistent comparison across all datasets, we convert literature SFR values to a \cite{2003PASP..115..763C} IMF using scaling factors of 0.63 and 0.94 for results based on \cite{1955ApJ...121..161S} and \cite{2001MNRAS.322..231K} IMFs, respectively. Our CSFRD estimates show good agreement with previous measurements from H$\alpha$ and UV tracers, and are slightly higher than the values predicted by the \cite{2014ARA&A..52..415M} relation, which adopted an integration limit of $0.03L_{*}$ similar to our approach. The dashed line in Figure \ref{fig: SFRD} corresponds to \cite{2006ApJ...651..142H} which adopted a significantly lower threshold of $\log[{\rm SFR}/(M_{\odot}~{\rm yr}^{-1})]=-4$, which shows similar trend compare with our results.
We also list relevant work in Figure \ref{fig: SFRD}, as well as the lower limits of SFR adopted for the CSFRD integration in Table \ref{tab: CSFRD_lower_limit}. 

Variations in this integration limit may lead to systematic bias in the CSFRD results. For the case of our parameters, integrating the star formation rate function down to the turnover luminosity of UV \citep[$M_{\rm UV}=-15.5$ mag,][]{2022ApJ...940...55B} would increase our CSFRD estimates by 9\%. Therefore, we estimate a 10\% systematic uncertainty in the integration limit across different methods and models.

In addition, there are discrepancies across methods. UV and H$\alpha$ trace a similar region to estimate the SFR, and their LF also shows a similar exponential decline at the bright end \citep{2009ApJ...692..778R, C2012, 2013MNRAS.428.1128S, 2017ApJ...838...29M}. However, the UV to H-alpha ratio still correlates with the stellar mass as discussed in Section \ref{subsec: Faint-end}. On the other hand, the SFR inferred by radio is based on the tight correlation observed between radio and FIR emission \citep{2001ApJ...554..803Y}, so the FIR and radio SFRs actually trace the dust emission. Their luminosity functions show a power-law decline at the bright end, less sharply than UV and H$\alpha$ results \citep{2011ApJ...730...61K, N2017,2018ApJ...853..172L, G2020, 2022ApJ...927..204E, 2024ApJS..275...36F,2023MNRAS.523.6082C,T2024}. This difference likely relates to the complex origins of dust emission, which is not solely produced in active star-forming regions \citep[e.g.][]{1987ApJ...314..513L,2009ApJ...700..161S}. Moreover, the detection limits of current FIR and radio surveys correspond to SFRs of $1-10~M_{\odot}\rm yr^{-1}$. Consequently, systematic uncertainties remain when comparing the faint-end constraints from H$\alpha$ and FIR/radio results. Finally, at higher redshifts, photo-z may introduce additional systematic biases in sample selection for methods that lack emission-line detection. It is worth noting that the IMF itself possibly varies with stellar mass \citep[e.g.,][]{2017MNRAS.468.3071N}, which could introduce an additional systematic bias into the inferred CSFRD.

Several systematic uncertainties remain in our CSFRD estimations. 
Firstly, our assumption of $A_{\rm H\alpha}=1.0$ mag is simplified and might be higher than actual values (e.g., \citealt{2013ApJ...779...34C} adopted $A_{\rm H\alpha}=0.5$). The $A_{\rm H\alpha}$ may also correlate with stellar mass and SFR, which could introduce additional uncertainty into the CSFRD.
Secondly, the conversion from $L_{\rm H\alpha}$ to SFR may have a bias for low-mass galaxies. As discussed in Section \ref{subsec: Faint-end}, the increasing $L_{\rm H\alpha}/L_{\rm UV}$ ratio in low-mass galaxies could lead to SFR overestimation. Moreover, studies of local dwarf galaxies \citep[e.g.,][]{2009ApJ...706..599L} indicate that H$\alpha$ may under-predict the total SFR compared to UV measurements. Finally, we set the lower integration limit of CSFRD to $0.4 M_{\odot}~\rm yr^{-1}$ (corresponding to our detection limit). 

\begin{deluxetable}{cc}
\tablecaption{\centering The lower limits of SFR adopted for the CSFRD integration
\label{tab: CSFRD_lower_limit}}
\tabletypesize{\footnotesize}
\tablewidth{80pt}
\tablehead{
\colhead{Work} & \colhead{SFR lower limit\tablenotemark{\tiny *}} \\
\colhead{} & \colhead{($M_{\odot}~\rm yr^{-1}$)}
}
\startdata
\multicolumn{2}{c}{$\rm H\alpha$-based} \\
\hline
This work & 0.40 \\
\cite{2013MNRAS.428.1128S} & \textbackslash \\
\hline
\multicolumn{2}{c}{UV-based} \\
\hline
\cite{2009ApJ...692..778R} & 0.37 \\
\cite{C2012} & 0.01 \\
\cite{2017ApJ...838...29M} & 0.01 \\
\hline
\multicolumn{2}{c}{IR-based} \\
\hline
\cite{2018ApJ...853..172L} & 0.01 \\
\cite{G2020} & 0.01 \\
\cite{2024ApJS..275...36F} & 1.00 \\
\cite{T2024} & 0.01 \\
\hline
\multicolumn{2}{c}{Radio-based} \\
\hline
\cite{2011ApJ...730...61K} & 0.02 \\
\cite{N2017} & \textbackslash \\
\cite{2022ApJ...927..204E} & 0.10 \\
\cite{2023MNRAS.523.6082C} & 2.57 \\
\enddata
\tablenotetext{*}{\textbackslash~corresponds to no lower limit of SFR in integrating the SFR density function.}
\end{deluxetable}

\subsection{AGN contamination}
The broad-line region of AGN can produce luminous H$\alpha$ emission, potentially contributing to the measured H$\alpha$ LF. It is therefore essential to evaluate the AGN contamination in our H$\alpha$ emitter sample and its impact on the following CSFRD estimation.

We find no broad-line components in the H$\alpha$ emitters of our sample, suggesting a negligible contribution from unobscured (Type 1) AGN with $\rm FWHM_{H\alpha}>2000~km~s^{-1}$. This can be explained by the relatively low volume density of broad-line AGN in the H$\alpha$ luminosity range probed here. We first cross-matched our sample with known AGN catalogs from MILLIQUAS \citep{2023OJAp....6E..49F}, SDSS \citep[DR17 \& DR18;][]{2022ApJS..259...35A, 2023ApJS..267...44A}, and DESI \citep[DR1;][]{2025arXiv250314745D}, which confirmed that none of our H$\alpha$ emitters are identified as known AGNs. We further estimated the expected number of AGN in our field using the QSO X-ray luminosity function from \cite{2020MNRAS.495.3252S}. By adopting the relations between $L_{5100}$ and $L_{\rm H\alpha}$ from \cite{2005ApJ...630..122G} and $L_{\rm bol}=10.33\times L_{5100}$ from \cite{2006ApJS..166..470R}, we extrapolated that the expected number of AGN in our survey volume (including $z\sim1.3$ and $z\sim2.0$) is approximately 0.3. This low value confirms that the probability of detecting an AGN in our limited field is indeed small.

We also considered the potential contribution from Type 2 AGNs, which are mostly identified via high-ionization emission lines or through mid-IR SED modeling. The NIRISS grism data cannot resolve the [N~\textsc{II}] lines required for some standard AGN diagnostics \citep[e.g., N2–Baldwin-Phillips-Terlevich diagram;][]{1981PASP...93....5B}. Furthermore, with HST and ground-based telescopes, it is hard to reach the depth in the UV or other short wavelengths necessary to match the sensitivity of JWST for our sources. 
The number densities of Type 1 and Type 2 AGNs are expected to be comparable at cosmic noon \citep{2006AJ....131.2766R, 2013ApJ...773...14R}, a view also recently supported by the observations of \cite{2025MNRAS.539.1562W}. Recent studies at low redshift indicate that approximately half of the H$\alpha$ emission in AGN host galaxies originates from star formation rather than AGN ionization \citep{2025arXiv251210186M, 2024MNRAS.535..123D}. Therefore, we conclude that the contribution of Type 2 AGNs to the observed H$\alpha$ luminosity density is negligible within the flux range in our study.

Finally, we considered the LRD contribution to our H$\alpha$ LF. While the number density of LRD at high redshifts may be ten times larger than traditional AGN, the majority of LRDs are found at z = 4 $\sim$ 8, regardless of whether the selection method is based on rest-frame properties or not \citep{2024ApJ...963..129M, 2024ApJ...968...38K, 2024arXiv241201887B, 2025ApJ...978...92L, 2025ApJ...986..126K, 2025ApJ...991...37A}. \citet{2025arXiv250408032M} found that the number density of LRDs at $z\sim 2$ is only about one-tenth that of quasars. Given the rapid decline in LRD number density with decreasing redshift, their contribution at $z\sim 1.3$ is expected to be negligible. Consequently, LRDs are unlikely to have a significant impact on our H$\alpha$ LF within the redshift range of this study.

\subsection{Methodological Implications}
Following the methodology of \cite{2023ApJ...953...53S} and \cite{ 2025ApJ...987..186F}, we estimate the observable volume and completeness individually for each emitter in our sample. Previous LF studies based on HST grism data primarily adopted two approaches. The first assumes an empirical functional for the relationship between overall completeness and observed flux, as introduced by \cite{1995AJ....109.1044F}. The parameters in such a relationship are determined through simultaneous fitting with the LF parameters \citep{2021ApJ...920...78B, 2023ApJ...943....5N}, or derived individually by assuming the intrinsic flux distribution of faint emission lines is a power law \citep{2019ApJ...875..152B}. In strong lensing fields, however, the spatial variation of magnification significantly increases the complexity of implementing these methods.

The second approach constructs a set of emission-line galaxy models with varying redshifts, brightnesses, spatial extents, and equivalent widths, placing them at different positions across the observed field, and then performs simulated extraction and visual inspection to obtain the completeness results \citep[e.g.,][]{2013ApJ...779...34C}. In comparison, our method is based on simulations for each identified H$\alpha$ emitter. For a sample of moderate size, our approach is more straightforward and, as a result, could avoid potential biases at the faint end introduced by extrapolating empirical relations.

Therefore, our method is well-suited for the observational configuration of JWST NIRISS deep fields targeting galaxy clusters. We plan to extend the application of this methodology to data from the CANUCS (PID 1208) and Technicolor (PID 3362) programs.

\section{Summary} \label{sec: summary}
In this work, we have identified H$\alpha$ emitters during cosmic noon in the ABELL-2744 field using NIRISS grism slitless spectroscopy from the GLASS-JWST program. Our results and conclusions are summarized as follows.

\begin{itemize}
    \item [1.] Using GRIZLI forward modeling, combined with visual inspection, we selected 99 H$\alpha$ emitters in the redshift range of z = 0.8-2.3. Thanks to the high sensitivity of JWST and the magnification provided by the strong lensing field, our sample reaches a depth approximately one order of magnitude deeper than previous 3D-HST and WISP surveys.
    \item [2.] We selected H$\alpha$ samples in two redshift bins, $z=1.05-1.5$ and $z=1.8-2.2$, corresponding to the prime wavelength coverages of the F150W and F200W filters, respectively. A novel method was developed to estimate the survey volume, completeness, and flux bias for each sample, quantifying the H$\alpha$ emitter number density.
    \item [3.] We measured the H$\alpha$ luminosity function at $z\sim1.3$ and $z\sim2.0$ down to luminosities of $L_{\rm H\alpha}\sim10^{40.5}~\rm erg~s^{-1}$ and $L_{\rm H\alpha}\sim10^{40.9}~\rm erg~s^{-1}$, respectively. These luminosities correspond to star formation rates of approximately 0.4 and 1.0 $M_{\odot}~\rm yr^{-1}$. By extending the constraints by $\sim$2 dex much fainter than the characterestic luminosity ($L_*$), we have for the first time robustly determined the faint-end slope $\alpha=-1.50$ at $z\sim1.3$ and confirmed the $\alpha=-1.61$ estimate at $z\sim2.0$ from previous ground-based narrow band surveys.
    \item [4.] We derived the integrated cosmic star formation rate densities (CSFRD): ${\rm CSFRD}=0.097^{+0.015}_{-0.016}~M_{\odot}~\rm yr^{-1}~Mpc^{-3}$ at $z\sim1.3$ and ${\rm CSFRD}=0.129^{+0.025}_{-0.030}~M_{\odot}~\rm yr^{-1}~Mpc^{-3}$ at $z\sim2.0$. These values are in agreement with previous dust-corrected UV and H$\alpha$ estimations but are slightly higher than those derived from infrared and radio surveys.
\end{itemize}

Our study demonstrates the efficiency and power of JWST NIRISS grism spectroscopy, especially when combined with gravitational lensing, for constructing complete emission-line galaxy samples. The two-dimensional grism data for our spectroscopically confirmed H$\alpha$ emitters, complemented by multi-band imaging in the cluster fields, will enable further detailed studies of galaxy properties. These include investigations into star formation histories, gas-phase metallicity gradients \citep[e.g.,][]{2022ApJ...938L..16W}, and the mass-metallicity relation \citep[MZR; e.g.,][]{matharu2025lookcompleteviewspatially}. The methodology presented here can be applied to other JWST NIRISS slitless spectroscopic programs, such as CANUCS \citep[PID 1208, PI: C. Willott;][]{2025arXiv250621685S}, TECHNICOLOR (PID 3362, PI: A. Muzzin), PASSAGE \citep[PID 1571, PI: M. Malkan;][]{malkanParallelApplicationSlitless2025a, 2021jwst.prop.1571M}, and OutThere (PID 3383 \& 4681, PI Glazebrook). Applying it will further our understanding of galaxy evolution at cosmic noon, from probing the bright end of the H$\alpha$ luminosity function in blank fields to searching for AGNs. The JWST data used in this paper are available from the GLASS-JWST High-Level Science Product collection at \href{https://archive.stsci.edu/hlsp/glass-jwst}{https://archive.stsci.edu/hlsp/glass-jwst}.

\acknowledgments
We thank Lei Sun, Hang Zhou for their useful discussions and suggestions. 
The authors would like to express their gratitude to the anonymous reviewer for his/her invaluable feedback and insightful suggestions.
This work is supported by the China Manned Space Program with grant no. CMS-CSST-2025-A06, the National Key R\&D Program of China No.2025YFF0510603, the National Natural Science Foundation of China (grant 12373009), the CAS Project for Young Scientists in Basic Research Grant No. YSBR-062, and the Fundamental Research Funds for the Central Universities. XW acknowledges the support by the Xiaomi Young Talents Program, and the work carried out, in part, at the Swinburne University of Technology, sponsored by the ACAMAR visiting fellowship. 
X.W. acknowledges the National Key R\&D Program of China (No. 2025YFA1614100).
MB acknowledges support from the ERC Grant FIRSTLIGHT and Slovenian national research agency ARIS through grants N1-0238 and P1-0188. BV and PW acknowledge support from the European Union – NextGenerationEU RFF M4C2 1.1 PRIN 2022 project 2022ZSL4BL INSIGHT and from the INAF Mini Grant ``1.05.24.07.01 RSN1: Spatially-Resolved Near-IR Emission of Intermediate-Redshift Jellyfish Galaxies'' (PI Watson).
This work is based on observations made with the NASA/ESA/CSA JWST, associated with program JWST-ERS-1324. We acknowledge financial support from NASA through grant JWST-ERS-1324. 
The JWST data presented in this article were obtained from the Mikulski Archive for Space Telescopes (MAST) at the Space Telescope Science Institute. The specific observations analyzed can be accessed via \dataset[DOI:10.17909/kw3c-n857]{https://doi.org/10.17909/kw3c-n857}.
The MAST at the Space Telescope Science Institute, which is operated by the Association of Universities for Research in Astronomy, Inc., under NASA contract NAS 5-03127 for JWST. 

\bibliography{GLASS_NIRISS_ELG}{}
\bibliographystyle{aasjournal}
\end{document}